\documentclass[12pt,preprint]{aastex}

\usepackage{graphicx}

\slugcomment{}
\shorttitle{A New Low-Mass Eclipsing Binary from SDSS-II}
\shortauthors{Blake et al.}

\begin{document}

\title{A New Low-Mass Eclipsing Binary from SDSS-II}
\author{Cullen H. Blake \altaffilmark{1}}
\affil{Harvard-Smithsonian Center for Astrophysics, 60 Garden Street,  Cambridge, MA 02138; cblake@cfa.harvard.edu}

\author{Guillermo Torres}
\affil{Harvard-Smithsonian Center for Astrophysics, 60 Garden Street,  Cambridge, MA 02138 }

\author{Joshua S. Bloom \altaffilmark{2}}
\affil{Astronomy Department, University of California, Berkeley, CA 94720 }

\author{B. Scott Gaudi}
\affil{Department of Astronomy, The Ohio State University, 140 W. 18th Ave., Columbus, OH 43210 }

\altaffiltext{1}{Harvard University Origins of Life Initiative Fellow}
\altaffiltext{2}{Sloan Research Fellow}

\begin{abstract}
We present observations of a new low-mass double-lined eclipsing binary system discovered using repeat observations of the celestial equator from the Sloan Digital Sky Survey II. Using near-infrared photometry and optical spectroscopy we have measured the properties of this short-period [$P$=0.407037(14) d] system and its two components. We find the following parameters for the two components: M$_{1}=0.272\pm0.020M_{\sun}$, R$_{1}=0.268\pm0.010R_{\sun}$, M$_{2}=0.240\pm0.022M_{\sun}$, R$_{2}=0.248\pm0.0090R_{\sun}$, $T_{\rm{1}}=3320\pm130$K, $T_{\rm{2}}=3300\pm130$K. The masses and radii of the two components of this system agree well with theoretical expectations based on models of low-mass stars, within the admittedly large errors. Future synoptic surveys like Pan-STARRS and LSST will produce a wealth of information about low-mass eclipsing systems and should make it possible, with an increased reliance on follow-up observations, to detect many systems with low-mass and sub-stellar companions. With the large numbers of objects for which these surveys will produce high-quality photometry, we suggest that it becomes possible to identify such systems even with sparse time sampling and a relatively small number of individual observations.

\end{abstract}

\section{Introduction}
Studies of  detached, double-lined eclipsing binary systems (DEBs) allow for direct measurements of the masses and radii of stars. These measurements have direct implications for theories of stellar structure and evolution. Our understanding of the structure of stars on the lower main sequence has improved over the past decade, but relatively few direct measurements of stellar mass and radius exist for objects with masses $0.1M_{\sun}<M_{\star}<0.4M_{\sun}$. Measurements of stars in this mass range allow us to test models of stellar structure where they are least well constrained by observation.  There are four known main sequence DEBs  inhabiting this mass range and the properties of these systems are well-measured (\citealt{ribas2003}, \citealt{metcalfe1996}, \citealt{hebb2006}, \citealt{vaccaro2007}). In order to refine physical models of low-mass stars, more systems must be discovered and studied. Large-area synoptic surveys offer the best hope for the identification of new low-mass eclipsing systems. The photometric databases of shallow, wide-angle, surveys such as those aimed at the discovery of GRB afterglows or transiting planets, are already being combed for the most interesting binary systems \citep{mercedes2006, devor2006}, but the number of M dwarfs in these shallow surveys is limited. These surveys have also been used to identify low-mass single-lined systems from which the mass and radius of an unseen companion can be inferred \citep{beatty2007}.

The advent of deeper surveys with sparser time sampling (less than one observation per day), such as Pan-STARRS \citep{hodapp2004} and LSST \citep{tyson2002},  will become a vital resource for studies of variable objects of all types. Since low-mass stars are intrinsically faint, deep surveys may prove particularly powerful for studies of these objects. The Sloan Digital Sky Survey (SDSS) and its follow-up survey SDSS-II are excellent prototypes of the types of data that these future surveys will produce. Here, we report the first results from a program to identify DEBs in SDSS-II observations of an area of the sky scanned repeatedly as part of  the SDSS  Supernova Survey program.  We present the discovery of a new DEB along with mass and radius measurements of these two low-mass stars based on spectra taken with Keck/LRIS and near infrared (NIR) photometry gathered with the robotic telescope PAIRITEL. This discovery helps to fill in a poorly sampled portion of the Mass-Radius relation for low-mass stars. In Section 2, we describe the identification of this DEB and the follow-up observations. In Section 3, we describe our analysis, modeling, and the derivation of the physical parameters of the system. In Sections 4 and 5 we discuss our results in the context of the known low-mass eclipsing binary systems and future prospects for the discovery of similar systems.

\section{Observations}

\subsection{Identification of SDSS-MEB-1}

The SDSS-II Supernova Survey \citep{frieman2008} repeatedly observes about 300 deg$^{2}$ of sky centered around the celestial equator. The raw data, both imaging and uncalibrated outputs from the photometric pipeline, are made publicly available through the SDSS SN Archive Server\footnote{\url{http://www.sdss.org/drsn1}}.  We downloaded all of the imaging data from $i$ band and the corresponding pipeline outputs  available as of April, 2006. In total, nearly $2\times10^{5}$ individual 2048 by 1489 pixel images from 70 observing runs were available. We calibrated the uncalibrated outputs of the photometric pipeline against stars from the SDSS DR4 \citep{adelman2006} photometric catalog.  This was done by matching the pixel coordinates of stars in the raw $i$-band images to the coordinates of preselected catalog F-G-K stars and calculating a simple photometric solution for each individual image using raw $PSF$ magnitudes (no color terms). Once the positions of stars in the raw $i$-band images were matched to the photometric pipeline outputs, a similar photometric solution was calculated in the $r-$ and $z$-bands.  Recent work by \citet{ivezic2007} and \citet{padmanabahn2007} presents more sophisticated methods for calibrating repeat observations in SDSS. We found our technique to produce individual light curves with RMS of between 0.03 and 0.05 mag ($i<$19.0),  adequate precision for the detection of large eclipse signals.

\begin{figure}
\begin{center}

\includegraphics*[angle=270, scale=.7, viewport=40 40 560 700]{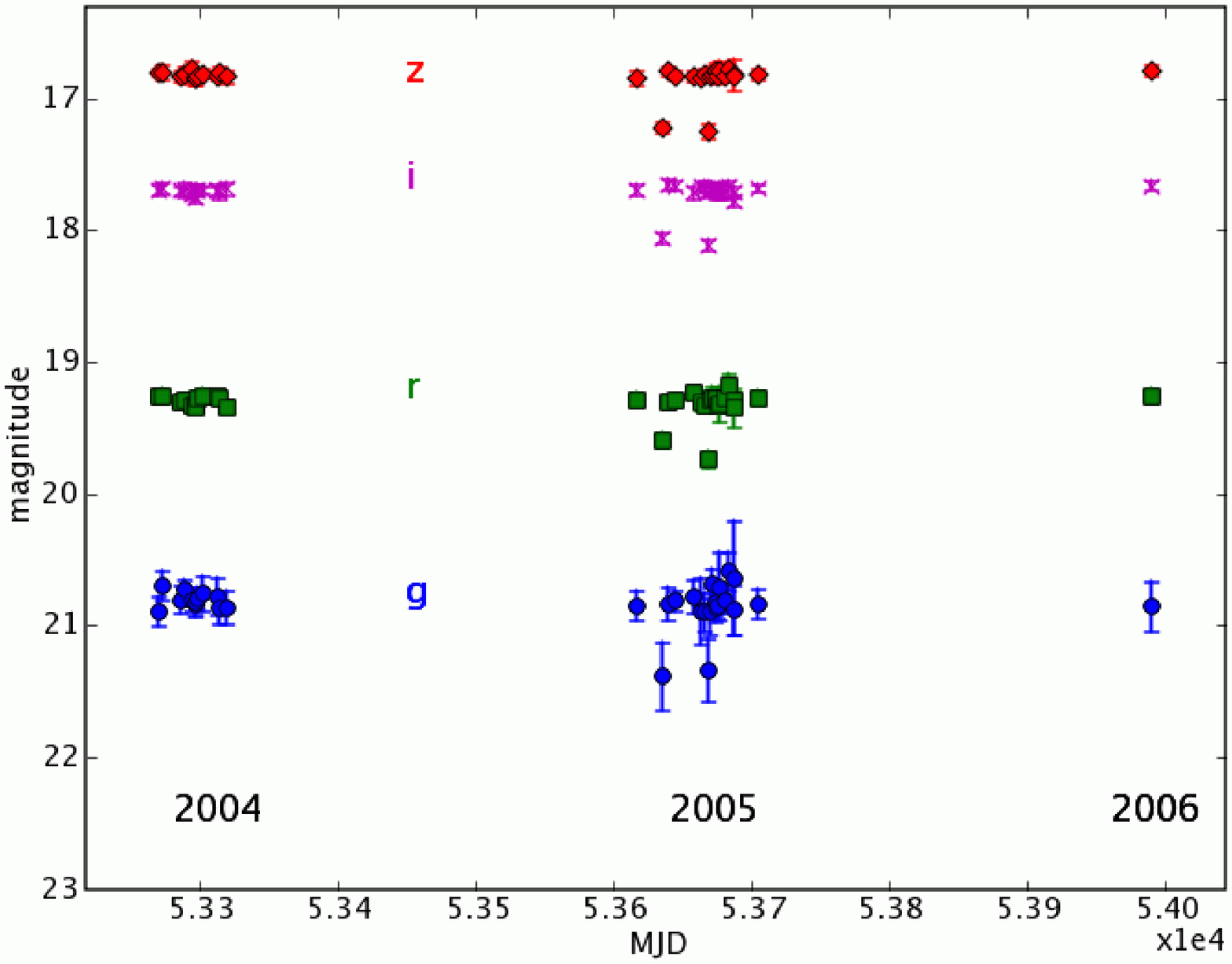}

\end{center}
\caption{ Multicolor light curve of SDSS-MEB-1 from the publicly available  
data in the first three seasons (2004 - 2006) of the SDSS II survey. Shown from top to 
bottom are the light curves in $z$-,  $i$-, $r$-, and $g$-bands. The 2005 observations during two eclipses are  
readily apparent in all bands. The zero-points of the images were  calibrated by tying the catalog PSF\_COUNTS fluxes to SDSS DR4 catalog  
magnitudes.}\label{figa}
\end{figure}

We preselected a set of 19,000 M dwarf targets from the  photometric database of SDSS DR4. This was done following M dwarf colors presented by \citet{west2005} and requiring that objects have flags corresponding to high photometric quality. Our targets all have $i-z>0.84$ and $r<21.2$, corresponding roughly to a sample of objects of spectral types M4 and later with $r$ band photometric precision better than $5\%$.  Faint objects with colors of $i-z>0.8$ are expected to be almost exclusively M dwarfs since other types of point sources, such as quasars and M giants, are very rare in this portion of the color-magnitude diagram \citep{fan1999}. We generated $r,i,z$ light curves for all of these objects, each light curve consisting of between 10 and 30 observations in each band. We identified potential eclipsing sources by searching for objects that exhibited decreases in brightness of more than 0.20 mag which were found to occur simultaneously in all three bands. This conservative requirement for the large amplitude of the dimming and the requirement that it be found in all three bands resulted in a robust method for identifying eclipsing candidates. Since we have relatively few observations of each target, we require a conservative detection threshold to ensure a very low rate of false positives. Our analysis produced a list of 13 candidates that have been observed as part of an ongoing program to identify DEBS in SDSS-II. Of these candidates, SDSS031824-010018 was the most promising. The SDSS-II light curve of this object, which we call SDSS-MEB-1, contained 15 points in $r,i,$ and $z$ and displayed two epochs with fluxes that were lower than the mean flux by $\Delta m>0.3$ mag in each band. An expanded SDSS-II light curve, containing additional data, is shown in Figure \ref{figa}. The basic observed parameters of SDSS-MEB-1 are given in Table 1. 

\subsection{Follow-up Photometry}

We observed SDSS-MEB-1 with the PAIRITEL robotic NIR observatory between 9 September, 2006 and 8 October, 2006. Observations consisted of 30 to 60 minutes of individual 7.8s integrations on each clear night. The PAIRITEL camera takes images in the $J$, $H$, and $K_{s}$ bands simultaneously. The PAIRITEL imager was originally used for the southern portion of the 2MASS survey \citep{skrutskie2006}. The observations were automatically scheduled,  collected, and reduced with the fully robotic observing system described in \citet{bloom2005} and \citet{blake2005}. The images were gathered using a randomized dither pattern and individual integrations were combined into mosaics with total integration time of $\sim$300s using \textit{SWarp} \citep{bertin2002}. We produced differential aperture photometry  relative to a set of comparison stars from the 2MASS catalog and photometric errors were estimated based on read noise and photon noise. Our analysis resulted in 937 individual measurements in $J$, $H$, and $K_{s}$ bands. We report the $J$, $H$, and $K_{s}$ measurements in Table 2.

\subsection{Spectroscopy}

We observed SDSS-MEB-1 spectroscopically using the LRIS \citep{oke1995} instrument on Keck on 20 September 2006  and 21 November 2006. Spectra were taken at resolutions R$\sim900$ and R$\sim3500$ using the $400$ grooves mm$^{-1}$, 8500 \AA ~blaze and $1200 $ grooves mm$^{-1}$, 7500 \AA ~blaze gratings with a 1$\arcsec$ slit. These two setups cover the spectral ranges 5550-9270 \AA ~and 6335-7600 \AA ~, respectively. We observed the standard star HD 19445 for flux calibration. We extracted the red channel spectra following the optimal extraction method outlined by \citet{horne1986}. We did not observe any radial velocity standards, so wavelength solutions were determined using both HgNeAr comparison lamps and isolated night sky emission lines from the catalog of \citet{osterbrock1997}. All spectra showed prominent emission due to $H\alpha$ and the $R\sim3500$ spectra revealed two distinct  emission lines. The detection of H$\alpha$ emission demonstrates that the stars have chromospheric activity. The percentage of low-mass stars that are found to be active peaks at spectral type M5, where up to $60\%$ of stars may show $H\alpha$ emission \citep{hawley1996}. We also obtained high-resolution imaging of the field around SDSS-MEB-1 on 15 October, 2006 using the SCAM camera on Keck/NIRSPEC. In $0.36\arcsec$ seeing we find no companions between 1$\arcsec$ and $25\arcsec$ down to an approximate limiting magnitude of $K$=18.5.

\section{Analysis}

An initial inspection of the PAIRITEL photometry revealed large changes in the observed brightness of this system, a clear signature of an eclipsing binary. We estimated the period of the system using the Phase Dispersion Minimization method of \citet{stellingwerf1978} in two stages. A first pass using coarse resolution in trial frequency indicated a prominent peak in the periodogram at around 0.4d with  lesser peaks at double and half this period. We re-calculated a periodogram around $P=0.4$d using a fine spacing in trial frequency of  2.5 day$^{-1}$.
We fit a parabola to the peak in this high-resolution periodogram to estimate the period of the system and found $P$=0.407037$\pm0.000014$d.
We determined a precise time of zero phase with a parabolic fit to the center of the primary eclipse in the folded light curve and estimated the epoch of primary eclipse to be $2453988.7993\pm0.0006$ (HJD). 

\begin{figure}
\epsscale{1.}
\plotone{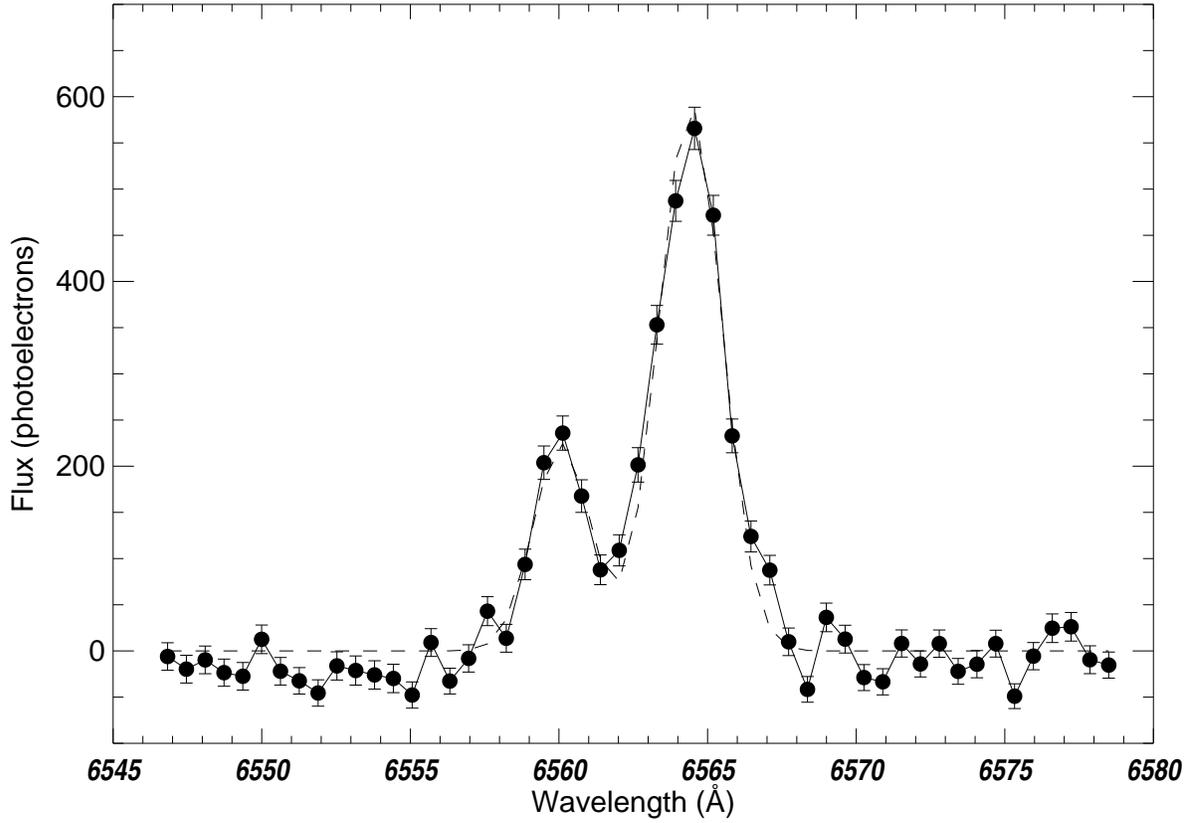}
\caption{Portion of the spectrum of SDSS-MEB-1 around the $H\alpha$ emission lines after continuum subtraction. The best-fit model is overplotted as a dashed line. This $R\sim3500$ spectrum was taken at HJD 2454060.978}\label{figb}
\end{figure}

We determined the radial velocities of both components of the binary by modeling the two $H\alpha$ emission lines seen in the $R\sim3500$ spectra. For both the $R\sim3500$ and $R\sim900$ spectra we began by fitting Gaussian profiles to isolated emission lines in comparison lamp spectra in order to determine the width of the instrumental profile. We found that the individual $H\alpha$ emission lines in the spectra of SDSS-MEB-1 were resolved in the $R\sim3500$ spectra so the Gaussian profiles were broadened to fit the data. We assume the same width for each of the $H\alpha$ emission lines. Two of our $R\sim3500$ spectra were acquired when the velocity separation between the two stars was close to maximal. In these spectra, the two emission lines are individually resolved and well separated and we fit a model of two super-imposed Gaussians to a $30\rm{\AA}$ portion of the spectra around $H\alpha$. This model has five parameters:  one for the centers of each Gaussian, one for the width of both Gaussians, one for their relative heights, and one for overall flux scaling.  We used $\chi^2$ minimization to fit for these parameters in the spectra taken near phase 0.25. A spectrum of SDSS-MEB-1 around the region of the two emission lines, along with the best-fit model, is shown in Figure \ref{figb}. We used the fits to these well-separated lines to fix the relative heights of the emission lines and their widths. The two emission lines are partially or totally blended in the rest of our data, particularly in the $R\sim900$ spectra where they are both unresolved and blended. For the remaining $R\sim3500$ spectra we used the fixed widths and relative heights determined previously to fit for the velocities of the two components. For the $R\sim900$ spectra we used the relative heights determined previously and widths corresponding to the instrumental profile determined from fits to emission lines in comparison lamp spectra.  We derived errors on the radial velocity estimates of each component in each spectrum based on the formal errors of the $\chi^2$ fit added in quadrature to the error estimate for the wavelength solution.  All of the radial velocities are corrected to the solar system barycenter using a the IRAF routine \textit{bcvcorr}. Our radial-velocity measurements are presented in Table 3 along with the estimated radial velocity curve parameters in Table 4. 

\begin{figure}
\epsscale{1.}
\plotone{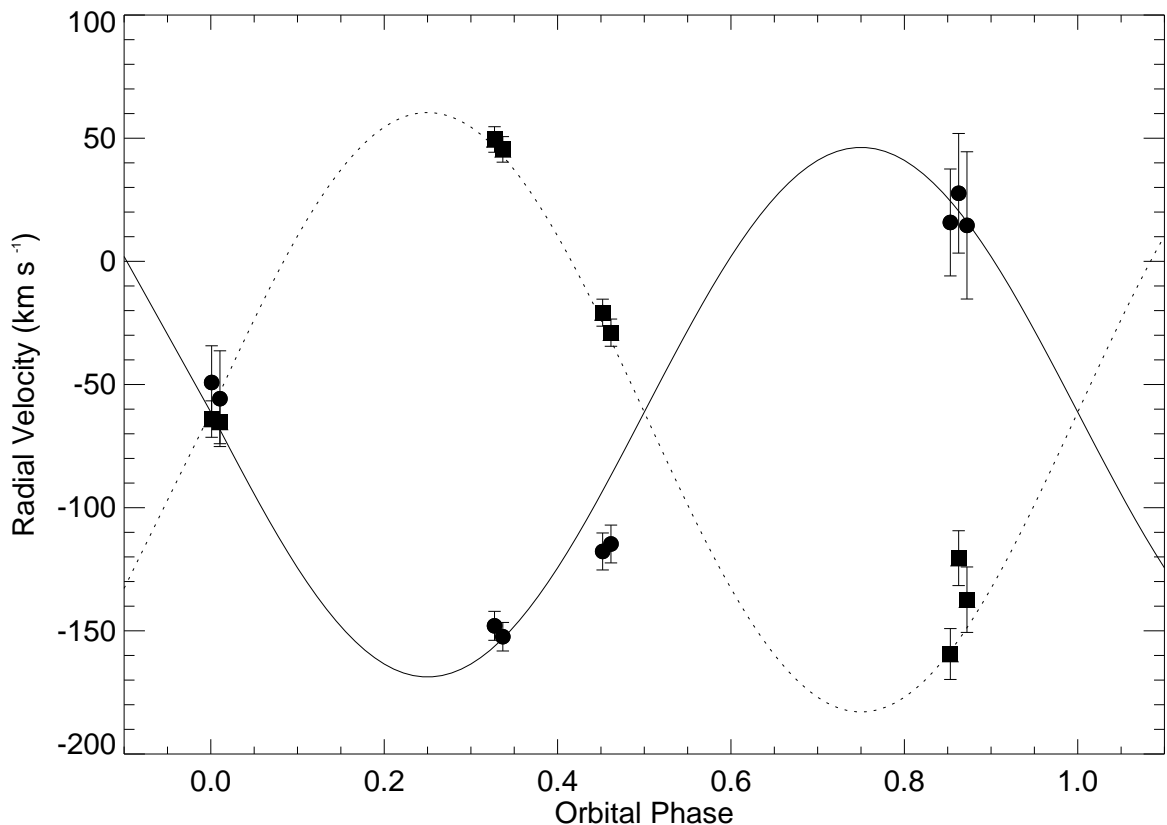}
\caption{Keck radial-velocity measurements of the two components of SDSS-MEB-1. The radial velocities are derived following the methods outlined in Section 3. For this fit, the period and epoch of primary eclipse estimated from the PAIRITEL photometry were assumed. A barycentric correction has been applied to these radial velocities. The squares correspond to the radial velocities of Star 2 and the circles correspond to the radial velocities of Star 1 ($M_{1} > M_{2}$). Star 2 eclipses Star 1 at phase 0.0 resulting in the slightly deeper eclipse depth seen at that phase in Figure 6.}\label{figd}
\end{figure}

While the uncertainty on our individual radial velocity measurements, particularly those from the $R\sim900$ spectra, are large, the radial velocities align very well with the epoch and period determined from the photometry alone. In order to estimate the system orbital parameters and the masses of each component we fixed the period and epoch of primary eclipse, assumed an eccentricity of $e=0.0$,  and fit a three parameter model to the radial velocity data. We estimated the two radial velocity amplitudes, $K_{1}$ and $K_{2}$, and the systemic velocity, $V_{\gamma}$ with a $\chi^2$ minimization. Assuming a Keplerian orbit with no eccentricity, the parameters of the radial velocity orbit translate easily into estimates for $M_{1} \sin^3i$, $M_{2}\sin^3i$, and $a\sin i$ \citep{book}. We estimated the errors on each of these values through a  bootstrap simulation.  We follow the convention that the primary star is the one eclipsed at phase 0, which is also the more massive star. This results in the marginally deeper eclipse depth seen at phase 0. The radial-velocity measurements, along with the best-fit model, are shown in Figure \ref{figd}.  Close inspection of the spectra resulting in two discrepant (off by $\sim 4 \sigma$ from the model) measurements from UT Nov 21 (near phase 0.45) indicated that the measurements were likely biased by the superposition of a sky emission line on the shoulder of the marginally-blended H$\alpha$ lines. In addition, some of our radial-velocity observations occur very close the time of primary eclipse. Assuming a rotation period locked to the orbital period of the system, we estimate that the amplitude of the Rossiter-McLaughlin effect is small comparable to our radial-velocity precision.

\begin{figure}
\plotone{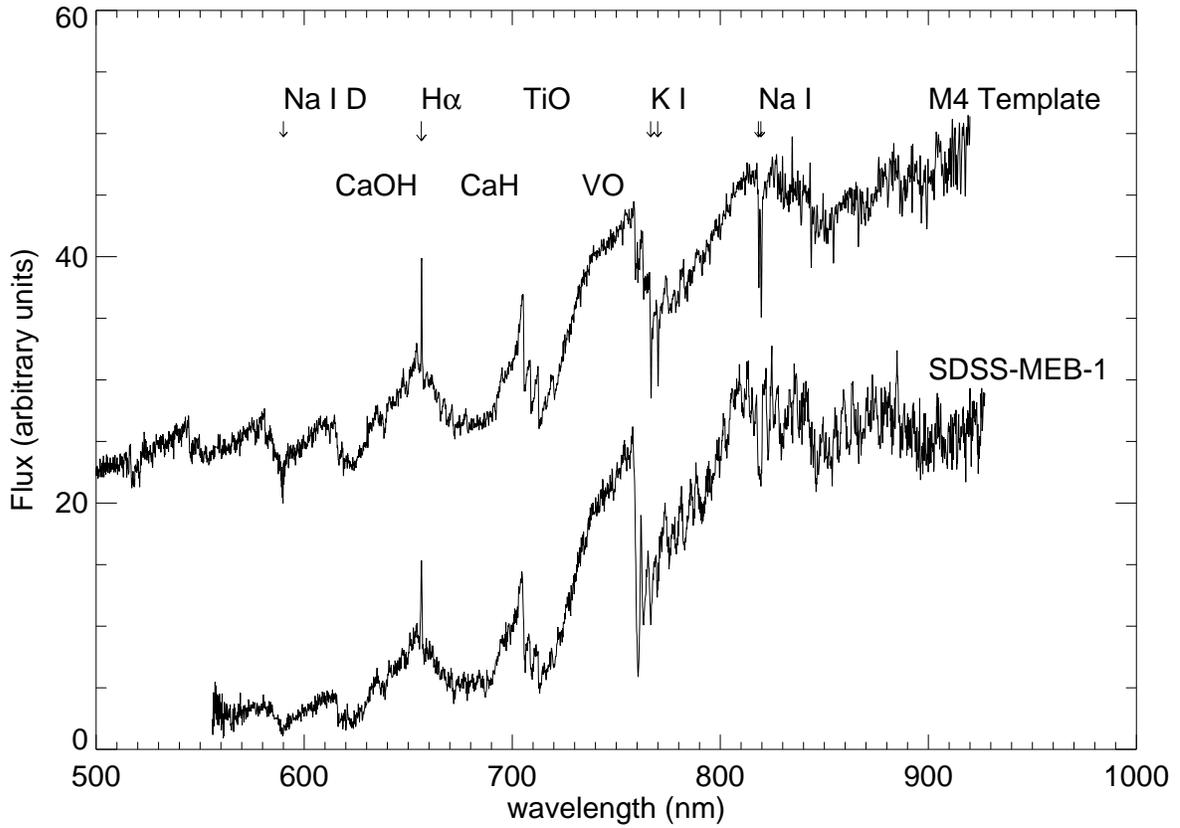}
\caption{Keck LRIS (R$\sim900$) Spectrum of SDSS-MEB-1 compared to the M4 template spectrum from \citet{bochanski2007}. Some broad molecular absorption features are indicated, along with atomic emission and absorption features (down arrows). }\label{figc}
\end{figure}

In order to increase the power of light curve modeling, and to break a partial degeneracy that exists in the determination of the ratio of the two stellar radii for systems with comparable eclipse depths or partial eclipses, we estimated the light ratio from our $R\sim3500$ spectra. We fit spectra taken near phase 0.0, when the component velocities are perpendicular to our line of sight, to template M dwarf spectra made available by \cite{bochanski2007}.  We found good agreement between our spectra and the active M4 dwarf template spectra. In Figure \ref{figc} we show the spectrum of SDSS-MEB-1 compared to the template of \citet{bochanski2007}. We solved for the light ratio by fitting spectra taken at other phases to the sum of two active M4 dwarf template spectra Doppler shifted relative to each other and each scaled by a different multiplicative factor. The Doppler shifts of the two template spectra were fixed to values defined by  the fit to the radial velocity curve. We calculated a $\chi^2$ goodness-of-fit between the model comprised of the weighted sum of the templates and our spectra for a range of relative brightnesses between 0.5 and 2.0. During the fitting, the area around the $H\alpha$ emission lines was given no weight. A fit to the region around the minimum of the resulting $\chi^2$ surface yielded a light ratio estimate of  $L_{2}/L_{1}=0.85\pm0.07$.  \citet{baraffe1998} presents theoretical evolutionary models of low-mass stars, the predictions of which we can compare to this estimate of the light ratio.   The spectra used in our analysis span the wavelength range $6400\rm{\AA}$ to $7540\rm{\AA}$, comparable to the $R$ band.  At an age of 2 Gyr and [M/H]=0.0, \citet{baraffe1998} predicts a light ratio of 0.744 (0.740 at t=5Gyr) in the Cousins/Bessel $R$ band for the stellar masses determined from the fit to the radial velocity curve. Based on the reasonable agreement between our light ratio estimate and the value predicted from theory, we adopt the NIR light ratio estimates of \citet{baraffe1998} throughout our modeling of the light curves. We estimated the error on this light-ratio estimate using a Monte-Carlo simulation by varying the masses (assuming $\sin i=1$) of the two components of the binary within their errors. We estimated an uncertainty of $\sim0.2$ in our light ratio estimate and do not include a systematic contribution due to any biases that may be inherent to the \citet{baraffe1998} models themselves. 

\begin{figure}
\epsscale{1.}
\plotone{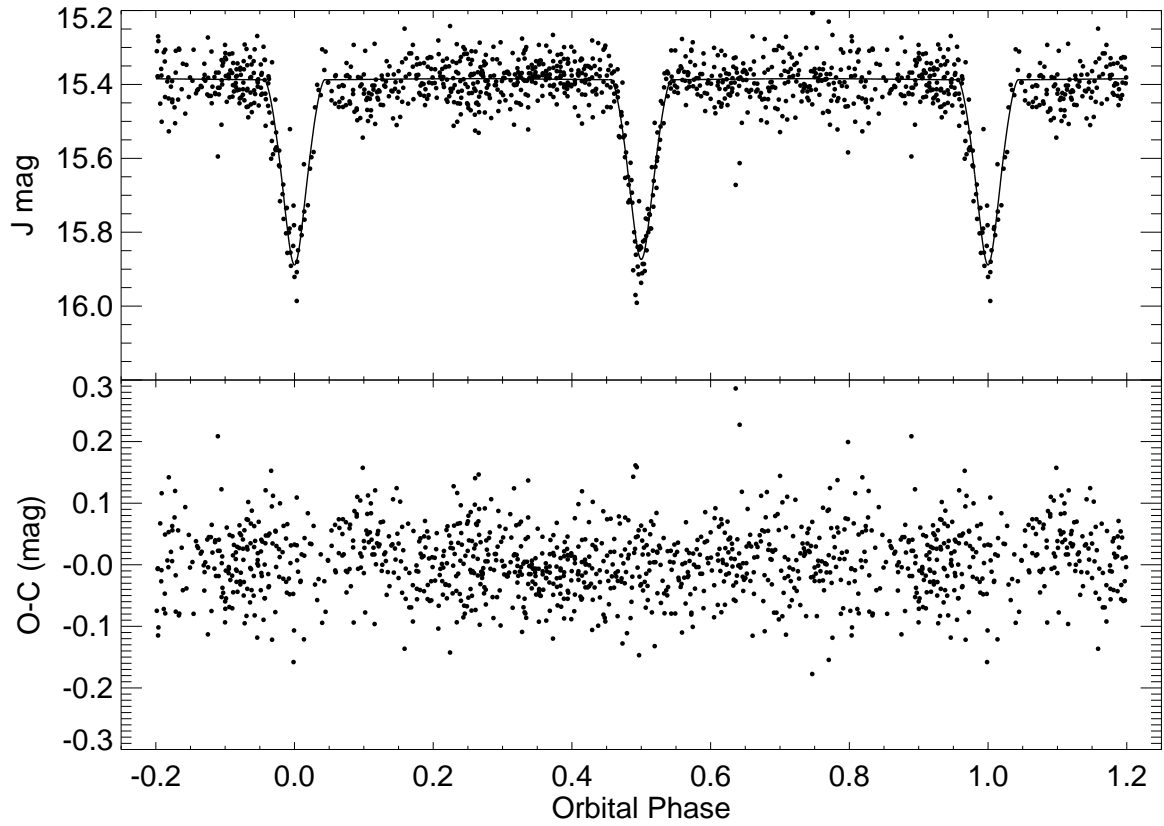}
\caption{\textit{Top:} PAIRITEL $J$-band differential light curve of SDSS-MEB-1. The best-fit model from an EBOP analysis of the light curve is overplotted. \textit{Bottom:} Residuals of the model fit to the data.}\label{fige}
\end{figure}

\begin{figure}
\epsscale{1.}
\plotone{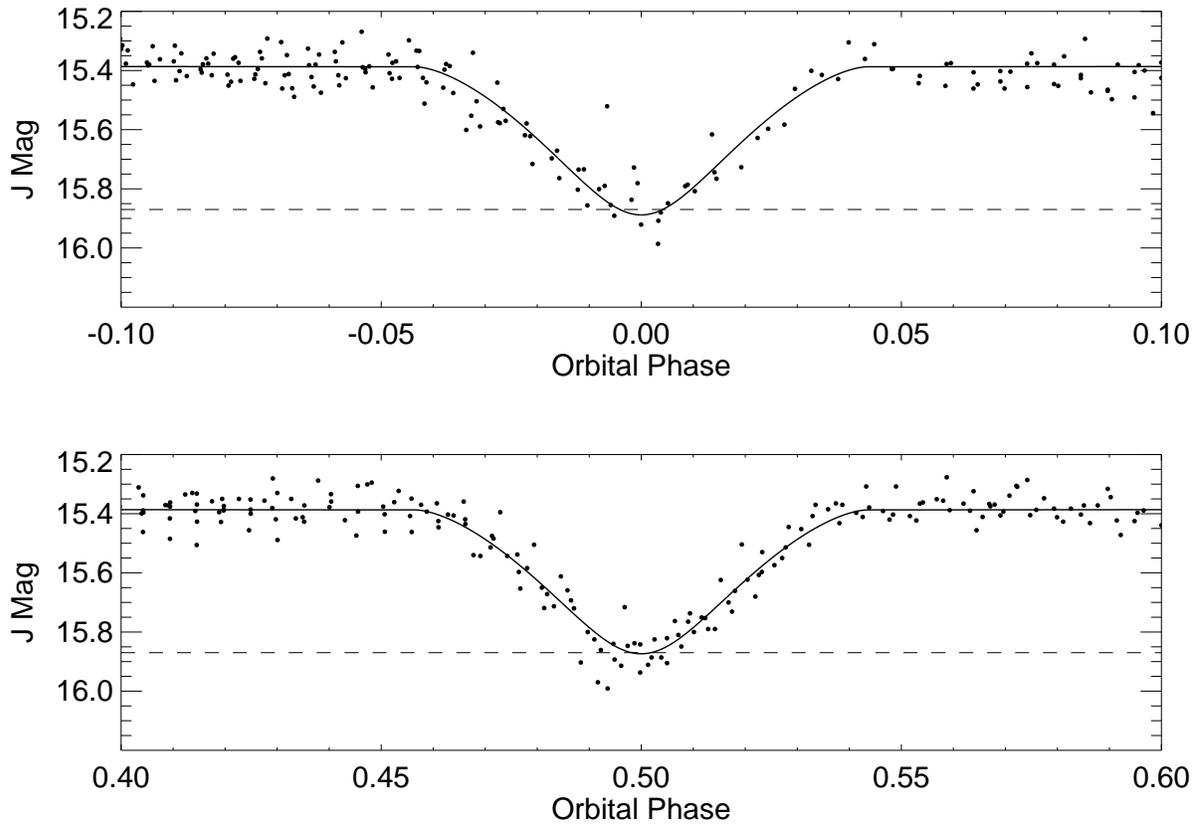}
\caption{Close up of the PAIRITEL $J$-band differential light curve around the two eclipses. The best-fit model from an EBOP analysis of the light curve is overplotted. The dashed line at $J$=15.87 is to guide the eye. The two eclipses are very similar in depth. Our best fit model is for a primary eclipse that is 0.015 mag deeper than the secondary.}\label{figf}
\end{figure}

We modeled our NIR light curves using the EBOP \citep{popper1981} technique as implemented in the code \textit{jktebop} by J. Southworth \citep{southworth2004}.  Here, we fixed the mass ratio, period, eccentricity, epoch, and the brightness ratio $L_{2}/L_{1}$ and fit for the sum of the radii of the stars in terms of the orbital separation, $(R_{1}+R_{2})/a$, the ratio of the radii of the stars, the inclination angle $i$, and the ratio of the central surfaces brightnesses, $J_{2}/J_{1}$.  We fixed the gravity darkening exponent to 0.32 as appropriate for a convective star \citep{lucy1967} and calculated the reflection coefficients from bolometric theory. We used logarithmic limb darkening coefficients for a $T_{\rm{eff}}$=3200K, $\log g =5.0$ star from \citet{claret2000} for NIR passbands. The \textit{jktebop} implementation of EBOP allows for the estimation of errors on the parameters using Monte-Carlo simulations. Based on the quality and quantity of the photometric data collected with PAIRITEL, as well as the long time baseline of the observations, we did not attempt to model spots on the components of the SDSS-MEB-1 system. Based on the EBOP models of the system, the out-of-eclipse variations are expected to be very small compared to the photometric errors. This allows us to evaluate the validity of our estimated photometric errors. We scaled our photometric error estimates until the reduced $\chi^{2}$ of the data at phase $0.7<\phi<0.9$, assuming constant brightness, was 1.0. We derived the errors on our estimated parameters from $10^4$ bootstrapping simulations and include our estimate of the uncertainty on the orbital separation $a$. We found other sources of error from fixed parameters, such as the mass ratio and the limb darkening model, to be negligible. We fit the $J-$, $H-$, and $K_{s}-$  band light curves with consistent results. The photometric errors on our $K_{s}$ photometry are large ($\sigma_{K_{s}}>0.15$)  compared to the eclipse depth and so the constraints on the system parameters from this band are relatively poor. We derived final values for the light curve parameters in two stages. First, the geometric ($i$, $R_{1}+R_{2}$, $R_{2}/R_{1}$) parameters were determined from weighted average of those from the $J$, $H$, and $K_{s}$ observations. With these geometric parameters fixed, we ran the EBOP analysis a second time to better determine the radiative parameters (central surface brightness ratio $J_{2}/J_{1}$ and total light ratio $L_{2}/L_{1}$). The model fit to our $J$-band data is shown in Figures \ref{fige} and \ref{figf} and the parameters from our EBOP analysis are shown in Table 5.  The physical parameters of the SDSS-MEB-1 system derived from the radial velocity and photometric analyses are given in Table 6. We estimate the ratio of the radius of each component to the radius of its Roche lobe to be 0.389 and 
0.367 for the primary and secondary, respectively.

We estimated the effective temperatures ($T_{\rm{eff}}$) of each component of SDSS-MEB-1 by combining the results of the EBOP analysis with the Barnes-Evans Color-Surface Brightness relation. The EBOP analysis of the NIR light curves provides $J_{2}/J_{1}$, the ratio of the central surface brightnesses of the two stars.  From the work of \citet{beuermann2006} we can relate the ratio of the average surface brightnesses of the two stars to the ratio of their $V-K$ colors.  Since the temperatures of the two stars are expected to be similar, our assumption of the same limb darkening model for each star allows us to assume that the ratio of the central surface brightnesses is the same as the ratio of the average surface brightnesses. Following the color-temperature relations of \citet{leggett1996} we convert NIR colors into temperatures. These relations are used to both turn the color ratio determined from the Barnes-Evans relation into a temperature ratio and to estimate the average temperature of the system.  From the combined color of the SDSS-MEB-1, we estimate the luminosity-weighted average temperature of the system.  Converting the DENIS $i_{D}$ to a Cousins $I_{c}$ following \citet{costado2005} and the 2MASS $K_{s}$ and $J$ to $K_{cit}$ and $J_{cit}$ following \citet{carpenter2001} we estimate the luminosity-weighted average temperature of the SDSS-MEB-1 system to be $3330\pm116 K$. The temperature ratio estimate and the estimate of the average temperature of the system comprise a system two equations in two unknowns, $T_{1}$ and $T_{2}$, which we solve numerically to arrive at the $T_{\rm{eff}}$ estimates given in Table 6. The errors on the temperature estimates are derived from Monte Carlo simulations in which the errors from each of the colors and color transformations are propagated.  We find that the temperatures of the two stars are consistent within the large errors and that the values agree well with the estimated temperatures for the components of the system CM Draconis, which have similar masses \citep{chabrier1995}. 

\section{Discussion}
\begin{figure}
\epsscale{1.}
\plotone{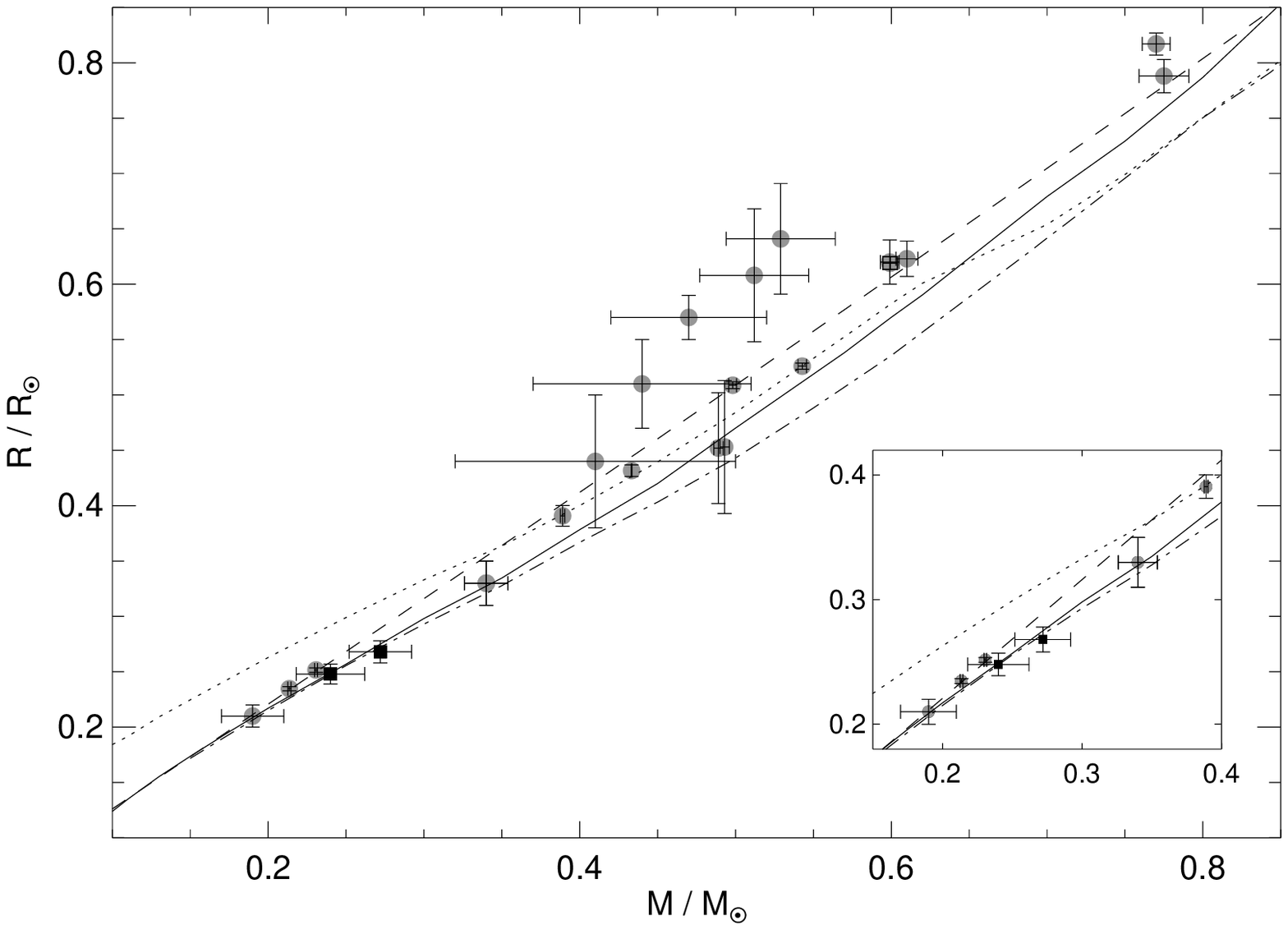}
\caption{Known low-mass stars with direct mass and radius measurements. The two components of SDSS-MEB-1 are plotted as black squares. Other objects taken from Table 10 of \citet{mercedes2005}, \citet{vaccaro2007}, \citet{hebb2006}, and the text of \citet{mercedes2006} are shown as gray circles. Overplotted are the empirical relation of \citet{bayless2006} (short-dashed line), a 5 Gyr, [M/H]=0 model from \citet{baraffe1998} (solid line), a 0.1 Gyr, [M/H]=0 model from \citet{baraffe1998} (dotted line) and a 5 Gyr, [M/H]=0 model from \citet{girardi2004} (dash-dot line).}\label{figg}
\end{figure}

\begin{figure}
\epsscale{1.}
\plotone{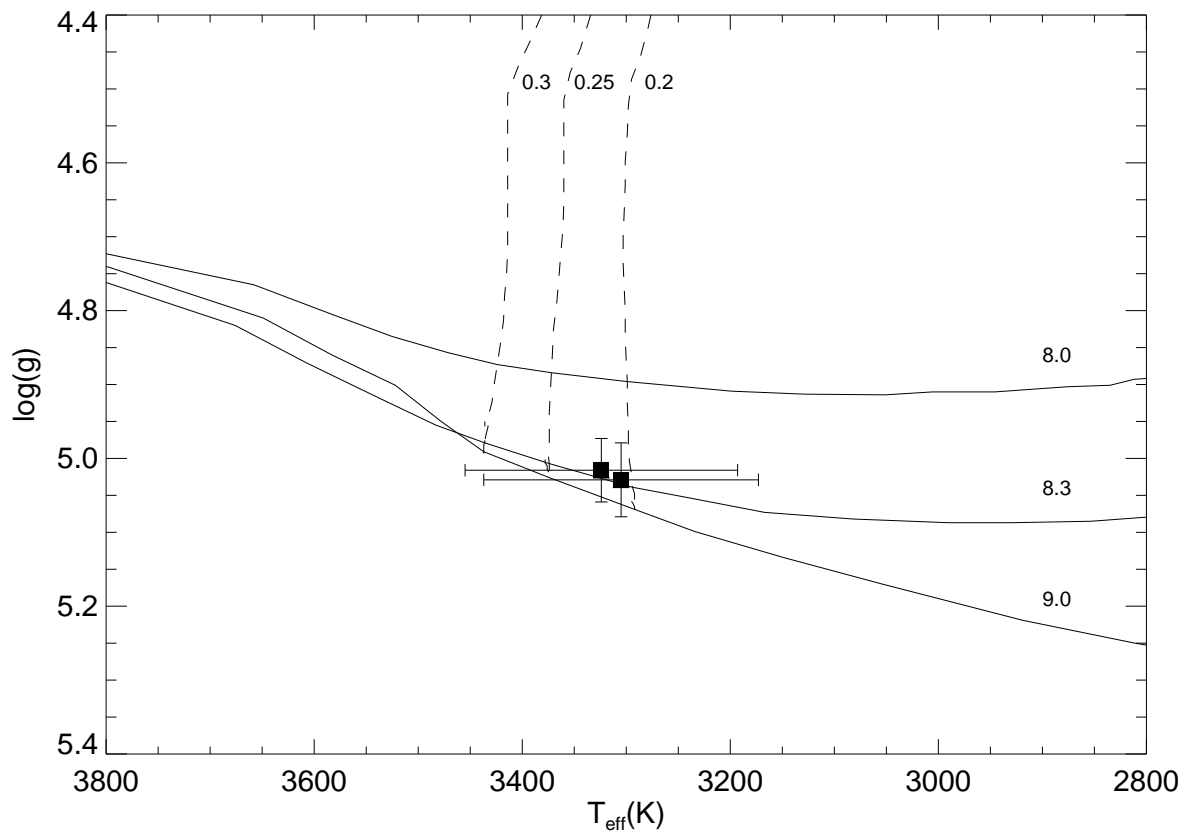}
\caption{The components of SDSS-MEB-1 compared to mass-tracks and isochrones from \citet{baraffe1998} models. The dashed lines are mass-tracks of masses 0.3, 0.25, and 0.2 $M_{\sun}$. The solid lines are [M/H]=0 isochrones of ages $\log t$=8.0, 8.3, and 9.0 years. From the combined light curve and radial velocity analyses, we estimate the masses of these two stars to be M$_{1}=0.272\pm0.020M_{\sun}$ and  M$_{2}=0.240\pm0.022M_{\sun}$}\label{figh}
\end{figure}

We combine the results of the light-curve and radial-velocity analyses to produce the physical parameters of the system listed in Table 6. The components of SDSS-MEB-1 fall in a mass regime were there are currently few direct mass and radius measurements from DEBs.  Our measurements are in good agreement with the empirical Mass-Radius relation of \citet{bayless2006} as well as the theoretical models ($\log t=$8.0, 9.0; [M/H]=0) of \citet{baraffe1998}. Recent work by \citet{chabrier2007} has explored the importance of rotation and magnetic activity  in the evolution of low-mass stars in binaries. Since both components of SDSS-MEB-1 show H$\alpha$ emission, demonstrating magnetic activity, a more precise set of system parameters could place important constraints on these new models. Our two new mass and radius measurements, along with models and other low-mass stars from the literature, are shown in Figure \ref{figg}. If our system were very young then our radius estimates would would fall about $20\%$ below the 0.1 Gyr relation of \citet{baraffe1998}. While SDSS-MEB-1 does lie in the vicinity of Orion, it is located more than $10^{\circ}$ from known star-forming regions. We also compare the $\log g$ and $T_{\rm{eff}}$ estimates from SDSS-MEB-1 to models from \citet{baraffe1998}. In Figure \ref{figh} we compare the two components of SDSS-MEB-1 to isochrones and mass-tracks. Using the solar metallicity models, we find a best-fit age for the system of $\log t  >8.1$. We find good agreement between the mass-tracks and isochrones from the \citet{baraffe1998} models. Both bias in the color-temperature relation we use and interstellar reddening could be important in temperature estimates of DEBs. According to the maps of \citet{sfd}, the total reddening in the direction of SDSS-MEB-1 in the NIR color we use to calibrate temperature is $E(I-J)=0.075$. Based on comparison of the observed $J$-$K$ colors of SDSS-MEB-1 to the same isochrones, we estimate a $J-$band modulus and a distance of the system of $\sim375$ pc. At the position of SDSS-MEB-1 relative to the Galactic disk, we expect a significant portion of the total estimated reddening along the line of sight. Since the average color of the system is an important factor in the determination of the individual temperatures, for $E(I-J)>0$ we will underestimate the temperatures of the components of SDSS-MEB-1. We estimate that for each 0.1 mag increase in $E(I-J)$ our estimated temperatures of the stars decreases by $85$K. 

With current data we lack sufficient precision to identify shortcomings in the models of low-mass stars. Increasing the total number of known low-mass DEB systems across a wide range of masses will improve our understanding of these objects. Deep, large-area surveys with sparse time sampling represent an excellent resource for the identification of new low-mass eclipsing systems. We have demonstrated that even as few as 15 high-quality, multi-color observations can be useful for identifying important new objects. While the amplitude of the eclipse signal is large enough to reliably identify promising candidates with synoptic surveys, objects discovered in this way present unique difficulties in terms of follow-up observations. The sparse sampling of the discovery light curves requires a significant investment of observing time to generate a complete light curve. Focusing on relatively faint targets with good photometric precision results in a vast increase in the total number of targets, particularly M dwarfs, and affords an increase in the number of viable targets for searching for eclipsing binaries and transiting planets. At the same time, the spectroscopic and photometric followup becomes more problematic and more expensive as the candidates become fainter. The study of eclipsing systems with future synoptic surveys will require a shift in emphasis from discovery to follow-up observations.  Robotic observing systems, such as PAIRITEL, are ideally suited for this type or work and will become an integral part of transient science produced by future synoptic surveys. 

\section{Future Prospects}
The discovery of SDSS-MEB-1 highlights the potential importance of the SDSS-II, and future synoptic
 data sets, for studies of eclipsing systems. The SDSS-II data are qualitatively 
different from the data sets typically used to search for low-mass eclipsing 
systems. Current planetary transit surveys tend to produce high-quality, high-cadence 
photometry for a relatively small number of targets. Future synoptic surveys will 
acquire fewer observations per target, with comparatively worse precision, but will 
observe orders of magnitude more targets.

We attempt to quantify the potential impact of SDSS-II on studies of low-mass stars by estimating the number of DEB systems that are likely to be present in the SDSS-II data. We used the $i-z$ to $i$-band absolute magnitude relations from \citet{west2005} and color relations of \citet{davenport2006} to estimate the absolute magnitudes of all of the point sources with $i-z>0.37$ contained in the SDSS-II repeat scan region. These absolute magnitudes were first converted to Bessel magnitudes and then to mass estimates following the solar-age, solar-metallicity isochrones of \citet{baraffe1998}.  Our estimate includes stars with mass $0.15M_{\sun}\le M \le 0.45M_{\sun}$. We assume a total binary fraction of $30\%$ \citep{delfosse2004} and integrate the mass-ratio distribution and semi-major axis distribution of \citet{duquennoy1991}  in order to estimate the total number of binaries in our survey and the distribution of their parameters.  We place several constraints on these binaries to estimate the number of systems that could be identified in the SDSS-II data and confirmed with follow-up observations. Since radial velocity observations are crucial for obtaining masses and radii of these stars, we constrain our targets to be brighter than $i=19.0$ with both components exhibiting radial-velocity semi-amplitudes of at least 30 km s$^{-1}$.  In order to ensure that the system is double-lined, we also require that the ratio of the luminosities of the components is no less than 0.1. We require that the system be in eclipse for more than $10\%$ of the time, so that the expected number of points in eclipse is $\sim2$ with SDSS-II, and that the amplitude of the eclipse is ${{\Delta F}\over{F}} > 10\%$. 

\begin{figure}
\epsscale{1.}
\plotone{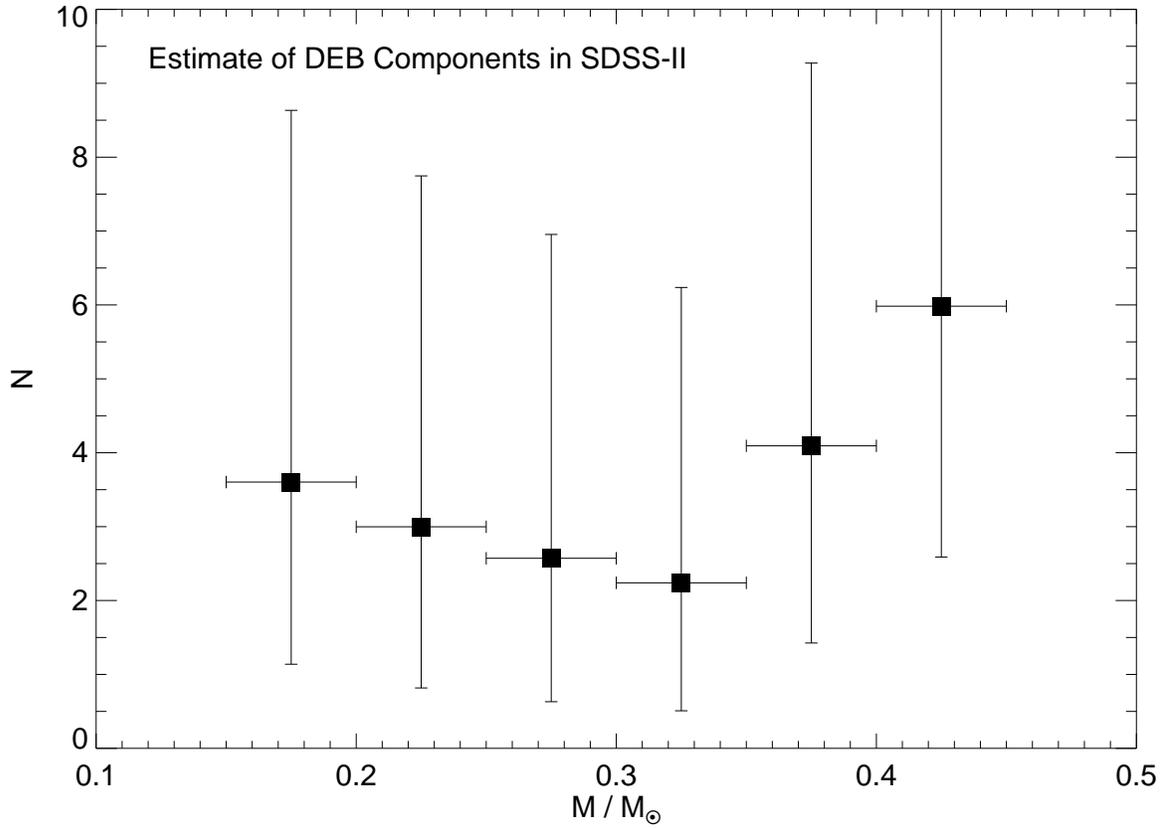}
\caption{Estimate of the total number of $0.15M_{\sun}<M<0.45M_{\sun}$ stars in DEB systems that can be identified in the SDSS-II repeat scan region and for which precise mass and radius determinations will be possible. The details of this simulation and the constraints placed on the targets are described in Section 4. Error bars correspond to $90\%$ confidence interval assuming Poisson statistics.}\label{figi}
\end{figure}

We estimate the total number of stars in systems meeting these criteria in mass bins of width 0.05 M$_{\sun}$ after scaling for the geometric probability of observing an eclipse for a given system. The results of our simulation are shown in Figure \ref{figi}. We estimate that the SDSS-II may contain as many as 12 low-mass eclipsing binary systems where both components are in the mass range $0.15M_{\sun}\le M \le 0.45M_{\sun}$. Our estimate attempts to count systems that are likely to be identified in the SDSS-II data and for which precise determination of mass and radius will be possible. Our simulation predicts the detection of between 1 and 7 stars ($90\%$ confidence) of mass $0.25M_{\sun}\le M \le 0.3M_{\sun}$, not inconsistent with our detection of the binary SDSS-MEB-1 (comprised of two low-mass stars). These estimates are only rough approximations since the binary fraction of M dwarfs, and the properties of their companions, are not well known.  Work by \citet{delfosse2004} points to an overabundance of short-period, $q\sim1.0$ M dwarf binary systems. Any such systematic differences in the properties of M dwarf and G dwarf binaries would effect the validity of our estimates.

The high quality, and shear quantity, of the data produced by synoptic surveys like SDSS-II also opens up the possibility of searching for small companions extending down into the regime of brown dwarfs and extrasolar planets.  
In principle, it is possible to detect eclipses of Jupiter-mass objects in the SDSS-II data. The giant planet and brown dwarf models of \citet{burrows2001} indicate that a companion more massive than $1M_{\rm{J}}$ results in an eclipse depth of ${{\Delta F}\over{F}} > 10\%$ for primaries with radii $R_{\star}<0.3R_{\sun}$. While the $10\%$ eclipse signal is probably at the limit of detectability with our current photometric calibration of the SDSS-II data, \citet{ivezic2007} have shown that homogeneous, $2\%$ photometry can be derived from repeat SDSS observations. Future work focusing on purely differential photometry of stars in the SDSS-II data could potentially extend sensitivity to Neptune-size planets orbiting the smallest stars.  Extrapolating from the 5 Gyr, solar-metallicity isochrones of \citet{baraffe1998} and the low-mass star colors of \citet{west2005} and \citet{davenport2006} (converting SDSS to Bessell colors) we estimate that our current selection criteria identifies $\sim$ 40,000 ($i<19.0$ ) targets with $R_{\star}<0.3R_{\sun}$. While detection of a transit by a Jupiter-mass object may be feasible, follow-up radial-velocity observations to confirm the mass of the system will be very difficult for all but the brightest objects. Our experience with SDSS-MEB-1 indicates that radial velocities with 10 km s$^{-1}$ precision should be feasible with reasonable exposure times for targets brighter than $i\sim19.0$. If we assume that a radial-velocity signal of semi-amplitude $K=30$ km s$^{-1}$ can be detected with a reasonable investment of telescope time, then companions with $M\sin i \ge 95 M_{\rm{J}}$ can be detected in orbits with periods $P\le 3$ days around all of these targets. \citet{blake2007} have demonstrated radial-velocity precision of 0.3 km s$^{-1}$ using high-resolution infrared spectroscopy, but only for relatively bright objects ($K<12.0$).  Unfortunately, only about 1100 of the M dwarfs we selected are this bright but for these targets we are capable of radial-velocity confirmation of companions as small as  $M\sin i \sim  3 M_{\rm{J}}$.  Still, radial velocity followup of these fainter systems may be feasible with the next generation of 30m ground-based optical telescopes.

Future deep, large-area synoptic surveys will contain a treasure trove of eclipsing systems and allow for the direct measurement of masses and radii of many low-mass objects down to the bottom of the main sequence. We demonstrate that photometry of faint objects with sparse time sampling can be used to discover important new eclipsing systems. With these systems, the challenge will lie in the follow-up observations necessary to produce precise estimates of the physical parameters of the system and it is likely that robotic observing facilities will play a crucial role in these investigations. We have presented here a new low-mass double lined eclipsing binary discovered in a data set that can serve as a prototype for the next generation of synoptic surveys, and demonstrated that follow-up observations of these objects are feasible. By extending to fainter objects and focusing on smaller stars it will be possible to greatly expand the number of viable targets for a search for eclipsing low-mass star and brown dwarf companions.

\acknowledgments
We would like to thank D. Charbonneau, L. Eyer, \v{Z}. Izevi\'{c}, D. Latham, and M. Wood-Vasey  for useful discussions and assistance that contributed to this work. We thank D. Starr for all of his work on the PAIRITEL system and J. Southworth for developing the \textit{jktebop} code and for making it available to the public. We also thank D. Perley and D. Kocevski for assistance with some of our Keck observations. We thank an anonymous referee for thoughtful comments that helped to improve this manuscript. CHB is supported by the Harvard Origins of Life Initiative. GT acknowledges partial support from NSF grant AST-0708229 and NASA's MASSIF SIM Key Project (BLF57-04). JSB is partially supported by Las Cumbres Observatory as well as the Sloan Research Foundation. The Peters Automated Infrared Imaging Telescope (PAIRITEL) is operated by the Smithsonian Astrophysical Observatory (SAO) and was   made possible by a grant from the Harvard University Milton Fund, the camera loan from  
the University of Virginia, and the continued support of the SAO and UC Berkeley.   The PAIRITEL
project is further supported by NASA/Swift Guest Investigator Grant   NNG06GH50G. We thank M. Skrutskie, E. Falco and the staff at Fred L. Whipple Observatory  for their continued support of the PAIRITEL project.  We wish to extend special thanks to those of Hawaiian ancestry on whose   sacred mountain we are privileged to be guests. Funding for the SDSS and SDSS-II has been provided by the Alfred P. Sloan Foundation, the Participating Institutions, the National Science Foundation, the U.S. Department of Energy, the National Aeronautics and Space Administration, the Japanese Monbukagakusho, the Max Planck Society, and the Higher Education Funding Council for England. 

{\it Facilities:} \facility{Keck/LRIS}

\clearpage

\clearpage

\clearpage

\clearpage

\clearpage

\clearpage

\clearpage

\clearpage

\clearpage

\clearpage

\begin{deluxetable}{lc}
\tabletypesize{\scriptsize}
\tablecaption{Basic Observables}
\tablewidth{0pt}
\tablehead{
\colhead{Parameter} & \colhead{Value} }
\startdata
RA (J2000) & 03:18:23.88\\ 
Dec (J2000) & $-$01:00:18.4\\
$g$ & 20.818$\pm0.041$\\
$r$ & 19.290$\pm0.015$\\
$i$ & 17.681$\pm0.007$\\
$z$ & 16.792$\pm0.010$\\
$J$ & 15.396$\pm0.048$\\
$H$ & 15.725$\pm0.056$\\
$K_{s}$ & 14.623$\pm0.101$\\
$i_{D}$ & 16.907$\pm0.100$\\

\enddata
\tablecomments{Position and brightness of SDSS-MEB-1. The $J$,$H$, and $K_{s}$ values are from 2MASS and the SDSS values are from the SDSS catalog \citep{adelman2006}. The $i_{D}$ magnitude is from the DENIS survey.}
\end{deluxetable}

\begin{deluxetable}{lccc}
\tabletypesize{\scriptsize}
\tablecaption{NIR Photometry}
\tablewidth{0pt}
\tablehead{
\colhead{HJD-2400000} & \colhead{Band} & \colhead{Mag} & \colhead{$\sigma$}} 
\startdata
 53995.8909  &  J  &  15.350  &  0.043  \\
 53995.8929  &  J  &  15.356  &  0.045  \\
 53995.8950  &  J  &  15.350  &  0.043  \\
 53995.8971  &  J  &  15.288  &  0.043  \\
 53995.8992  &  J  &  15.422  &  0.067  \\
 53995.9013  &  J  &  15.295  &  0.045  \\
 53995.9034  &  J  &  15.323  &  0.064  \\
 53995.9056  &  J  &  15.393  &  0.048  \\
 53995.9077  &  J  &  15.406  &  0.048  \\
 53995.9098  &  J  &  15.543  &  0.051  \\
..... & ..... & .....\\
\enddata
\tablecomments{PAIRITEL NIR photometry of SDSS-MEB-1. Errors reported here have been scaled as described in Section 3. Only a portion of the photometry is reported here. All of the data are available in the on-line edition. }
\end{deluxetable}

\begin{deluxetable}{lccccc}
\tabletypesize{\scriptsize}
\tablecaption{Radial-velocity Measurements}
\tablewidth{0pt}
\tablehead{
\colhead{HJD-2400000} & \colhead{RV$_{1}$} & \colhead{$\sigma_{RV_{1}}$}  & \colhead{RV$_{2}$} & \colhead{$\sigma_{RV_{2}}$} & \colhead{setup}\\
\colhead{} & \colhead{km s$^{-1}$} & \colhead{km s$^{-1}$} & \colhead{km s$^{-1}$} & \colhead{km s$^{-1}$} & \colhead{}}
\startdata
54000.13646  & 15.79  &  21.70 &  $-$159.32  &  10.33  &  400/8500\\
54000.14036  & 27.63  &  24.28 &  $-$120.50  &  11.14  &  400/8500\\
54000.14427  & 14.60 &  29.89 &  $-$137.36  &  13.29  &  400/8500\\
54060.84514  & $-$49.19  &  14.90 &  $-$64.03  &  7.39  &  1200/7500\\
54060.84904  &  $-$55.76  &  19.44 &  $-$65.21 &  8.83  & 1200/7500\\
54060.97794  &   $-$147.97  &   5.88 & 49.48  &   5.19  &  1200/7500\\
54060.98185  &   $-$152.38  &   5.79 & 45.47  &   5.18  & 1200/7500\\
54061.02872  &  $-$117.77  &  7.52 & $-$20.83  &   5.50  & 1200/7500\\
54061.03263  &  $-$114.74  &  7.67 & $-$28.96  &   5.52  &  1200/7500\\

\enddata
\tablecomments{Radial-velocity measurements of the two components of SDSS-MEB-1. These values were estimated following the procedures outlined in Section 3. The last column indicates the groves mm$^{-1}$ (400 or 1200) and the blaze angle ($7500\rm{\AA}$ or $8500\rm{\AA}$) of the grating used.}
\end{deluxetable}

\begin{deluxetable}{lc}
\tabletypesize{\scriptsize}
\tablecaption{Radial Velocity Fit Parameters}
\tablewidth{0pt}
\tablehead{
\colhead{Parameter} & \colhead{Value}} 
\startdata

V$_{\gamma}$ (km s$^{-1}$)  &  $-$61.24$\pm2.31$\\
K$_{1}$ (km s$^{-1}$) & 107.51$\pm5.22$\\
K$_{2}$ (km s$^{-1}$) & 121.70$\pm4.22$\\
O$-$C rms (km s$^{-1}$) & 13.96\\
e (fixed) & 0.0 \\

\enddata
\tablecomments{System parameters for SDSS-MEB-1 estimated from modeling  spectroscopic observations  as described in Sections 2 and 3. }
\end{deluxetable}

\begin{deluxetable}{lcccc}
\tabletypesize{\scriptsize}
\tablecaption{Light-Curve Fit Results}
\tablewidth{0pt}
\tablehead{
\colhead{Parameter} & \colhead{$J$ Band}  & \colhead{$H$ Band } & \colhead{K$_{s}$ Band} & \colhead{Adopted Value}} 
\startdata

($R_{1}+R_{2}$)/a & 0.279$\pm{0.006}$ & 0.284$\pm{0.008}$ & 0.270$\pm{0.022}$ & 0.280$\pm{0.005}$\\

$R_{2}/R_{1}$ & 0.928$\pm{0.019}$  & 0.911$\pm{0.022}$ & 0.870$\pm{0.045}$ & 0.916$\pm{0.014}$\\

$R_{1}/a$ & 0.144$\pm{0.0034}$ & 0.149$\pm{0.0050}$ & 0.140$\pm{0.012}$ & 0.145$\pm{0.003}$\\

$R_{2}/a$ & 0.134$\pm{0.0034}$ & 0.135$\pm{0.0039}$ & 0.126$\pm{0.012}$ & 0.134$\pm{0.003}$\\

i ($^{\circ}$) & 85.87$\pm{0.59}$ & 85.56$\pm{0.35}$ & 85.72$\pm{0.73}$ & 85.65$\pm{0.28}$\\
\\

$\rm{J}_{2}/\rm{J}_{1}$ & 0.977$\pm{0.029}$ & 0.980$\pm{0.048}$ & 1.075$\pm{0.100}$ \\
$L_{2}/L_{1}$ (input) & 0.792$\pm{0.230}$ & 0.811$\pm{0.200}$ & 0.818$\pm{0.200}$\\
$L_{2}/L_{1}$ (best-fit) & 0.842 & 0.814 & 0.813\\
LD [c,d] (fixed) & [0.5619,0.5212] & [0.5471,0.5310] & [0.4595,0.4603] & \\

\\
$P$ (day) & & & & 0.407037$\pm0.000014$\\
$T_{\rm min}$ (HJD) & 	& & &  2453988.7993$\pm0.0006$\\
O-C rms (mag) & 0.06 & 0.08 & 0.14\\

\enddata
\tablecomments{System parameters for SDSS-MEB-1 estimated from modeling of the photometric data. Limb darkening (LD [c,d]) parameters are for the logarithmic model of \citet{claret2000}. The value of $J_{2}$ is the relative central surface  brightness of the secondary ($J_{1}=1.0$). $T_{min}$ is the time when the primary is eclipsed. The $O-C$ values indicate the rms of the residuals of the data minus the model.}

\end{deluxetable}

\begin{deluxetable}{lc}
\tabletypesize{\scriptsize}
\tablecaption{Physical Parameters for SDSS-MEB-1}
\tablewidth{0pt}
\tablehead{
\colhead{Parameter} & \colhead{Value}} 
\startdata

$M_{1}$ (M$_{\sun}$) & 0.272$\pm0.020$\\
$M_{2}$  (M$_{\sun}$) & 0.240$\pm0.022$\\
$a (R_{\sun}$) & 1.850$\pm0.047$\\
$R_{1}$ (R$_{\sun}$) & 0.268$\pm0.0090$\\
$R_{2}$ (R$_{\sun}$) & 0.248$\pm0.0084$\\
$\log g_{1}$ & 5.016$\pm0.043$\\
$\log g_{2}$ & 5.029$\pm0.050$\\
$T_{1}$ & $3320\pm130$ K\\
$T_{2}$ & $3300\pm130$ K\\
M$_{\rm{V}}$$_{1}$ & $11.55\pm0.30$\\
M$_{\rm{V}}$$_{2}$ & $11.86\pm0.37$\\

\enddata
\tablecomments{System parameters for SDSS-MEB-1 estimated from modeling of the photometric and spectroscopic followup described in Sections 2 and 3. The values for surface gravity, $\log g$, are given in $cgs$ units. The estimates of the absolute
 $V$-band magnitudes of the stars are based on fits to \citet{baraffe1998} models }
\end{deluxetable}

\end{document}